\documentclass[conference]{IEEEtran}
\usepackage{amsmath,amsthm}
\usepackage{cite}
\usepackage{graphicx}
\usepackage{epstopdf}
\usepackage{amsfonts,amsmath,amssymb}
\usepackage{cite}
\usepackage{graphicx}
\usepackage{bm}
\usepackage{bbm}
\usepackage{amssymb}

\begin{document}

\title{Quantum Geo-Encryption}

%\author{Robert~Malaney$^*$
%%\emph{Member IEEE}
%       % <-this % stops a space
%%\thanks{}% <-this % stops a space
%
%\thanks{$^*$Robert Malaney is with the School of
%Electrical Engineering  and Telecommunications at the University
%of New South Wales, Sydney, NSW 2052, Australia. email:
%r.malaney@unsw.edu.au} }

\author{
\IEEEauthorblockN{ Robert Malaney}
\IEEEauthorblockA{School of Electrical Engineering  \& Telecommunications,\\
The University of New South Wales,\\
Sydney, NSW 2052, Australia\\
r.malaney@unsw.edu.au}
}

\vspace{-3cm}

\maketitle
\begin{abstract}

In this work we introduce the concept of \emph{quantum geo-encryption} - a protocol that invokes  direct quantum encryption of messages coupled to quantum location monitoring of the intended receiver. By obfuscating the quantum information required by both the decrypting process and the  location verification process, a communication channel is created in which the encrypted data can only be \emph{decrypted} at a specific geographic locale. Classical wireless communications can be invoked to unlock the quantum encryption process thereby allowing for any deployment scenario regardless of the channel conditions. Quantum geo-encryption can also be used to realize quantum-computing instructions that can only be implemented at a specific location, and allow for a specified geographical data-route through a distributed network.  Here we consider the operational aspects of quantum geo-encryption in  generic Rician channels, demonstrating that the likelihood of a successful spoofing attack approaches zero as the adversary moves away from the allowed decrypting location.
The work introduced here resolves a long-standing quest to directly deliver information which can only be decrypted at a given location free of assumptions on the physical security of a receiver.

\end{abstract}

%\section{Introduction}

 %Location information plays an ever-increasing role in modern telecommunication systems.

\section{Introduction }
 Geo-encyryption is an enhancement to traditional security techniques in which the decryption process can only occur at a specific geographic location. Clearly such an outcome, if possible, would have important implications for a wide range of  scenarios. Many attempts have been made to deliver a viable geo-encryption process, with varying degrees of success \cite{1,2,3,4,6,7,8,10,11}. However, the security of all known attempts at geo-encryption is beholden to some assumed restriction on the ability of an adversary, and/or tamper-proof assumptions on a specific device. Ultimately, the necessity of such assumptions detract from the geo-encryption paradigm and commercial implementations of it have not  occurred.

 In this work we remove all debilitating assumptions from the geo-encryption paradigm by introducing quantum information into the decryption process. More specifically, we  show how the use of quantum information embedded on a communication device can be used to simultaneously invoke  quantum direct communication (QDC) and quantum location verification (QLV) that collectively realize a geo-encryption solution that is secure against any known attack. We further show how this new form of geo-encryption can be extended to make the decryption of data viable not only at a specific location, but also a specific time. % We explore the use of this new form of geo-encryption in the most generic of wireless communication channels - the Rician channel.
 %We also discuss the usefulness of quantum geo-encryption in allowing for quantum computations viable only at a specific quantum computer at a specific time. Implications of this new  application for distributed quantum computation and secured information routing paths are also discussed.

%\section{Related Work}

The notion of geo-encryption first appeared in the scientific literature in 1996 where geo-encryption based on GPS signals was discussed \cite{1}. The main idea behind such classical  geo-encryption is the addition of a so-called \emph{geo-lock} to the standard SSL (Secure Sockets Layer) protocol \cite{2}. The main idea can be summarized thus.
\emph{Encryption:-}
(i) A message is encrypted in the normal manner by the sender using a conventional cipher and associated session key $R_k$, randomly generated by the sender. (ii) The sender  decides  the location $L_d$ and time $t_d$ for which decryption is to be allowed. (iii) Based on this $L_d$ and $t_d$, the sender creates a \emph{geo-lock} $G_L$, which is a binary string  determined using an (known) algorithm to convert information related to the anticipated signal metrics a receiver at the decryption location and time should obtain (e.g. specific frequencies at specific signal strengths). (iv) The sender uses the geo-lock $G_L$ to XoR the key $R_k$, forming a new key $G_k$. This new key is then encrypted using a standard  Rivest-Shamir-Adleman (RSA)   private key  to form another new key, $G_E$.
\emph{Decryption:-}
(v) The receiver captures the  encoded message and the key $G_E$
using his own radio antenna(s). (vi) The receiver proceeds to decrypt $G_E$ using the sender's RSA public key to obtain $G_k$.
(vii) From the received signal metrics the receiver computes $G_L$, and applies that to $G_k$ to compute the session key $R_k$.
(viii) The original message is decrypted.  A schematic of this process is illustrated in Fig.~1.

Pivotal to the success of this protocol is that the received signal information associated with the geo-lock cannot be spoofed. Indeed, a series of steps can be put in place in order to minimize such spoofing in civilian receivers \cite{2}, and in the military sphere
  anti-spoof GPS receivers based on encoded GPS signals are available. However, such techniques are not immune to device tampering and/or the leaking (or determination) of GPS decoding keys. This is particularly so, when a quantum adversary is in place - namely an adversary whose resources and abilities are confined only by the laws of physics. In the presence of such an all-powerful adversary, the above classical geo-encryption protocol is readily attacked. In this work we will assume the presence of a quantum  adversary.

\begin{figure}[!t]
    \begin{center}
    {\includegraphics[width=3.4 in]{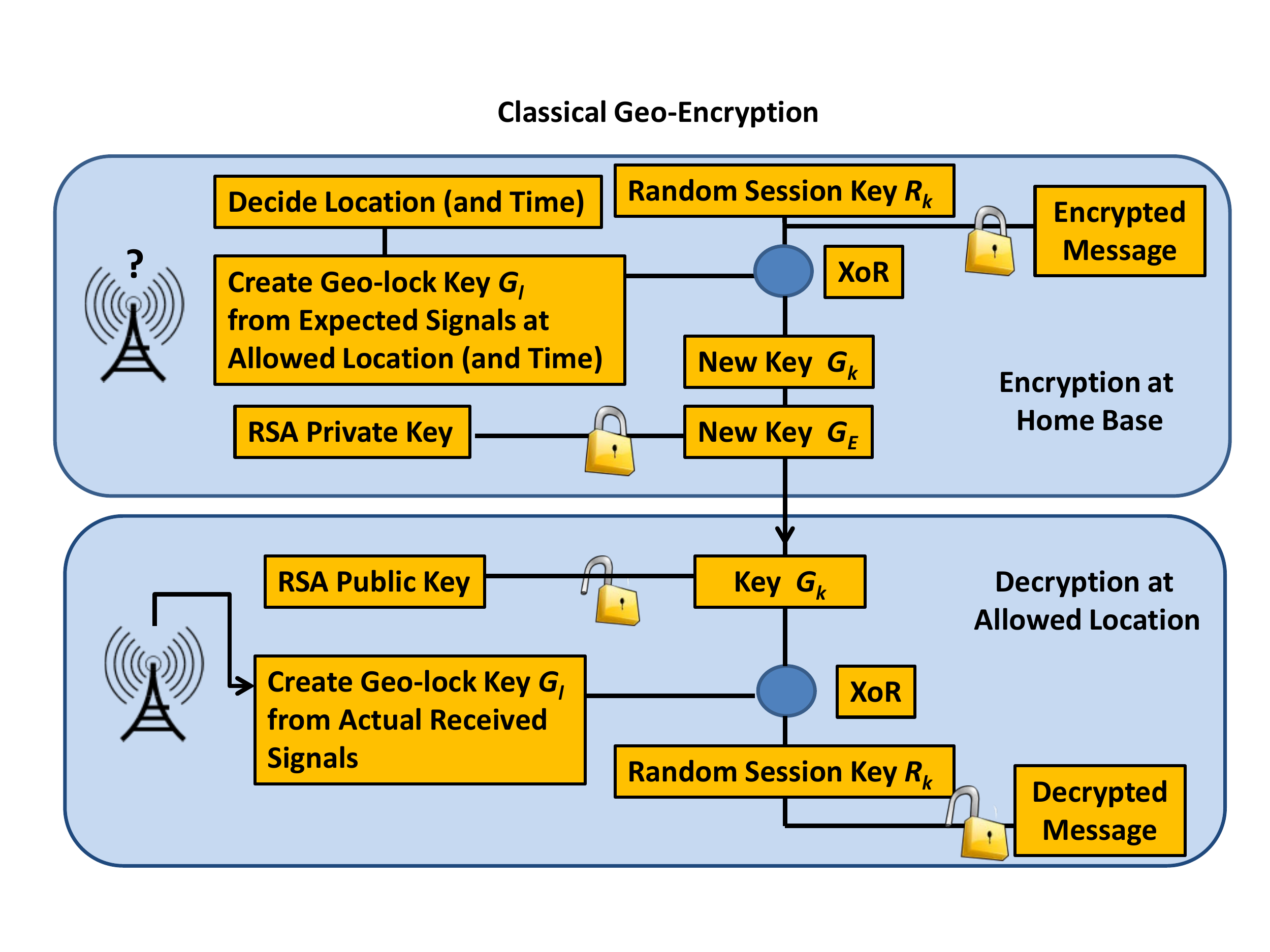}}
    \caption{Classical Geo-Encryption. Although many details can be different in different versions of this protocol, the use of a `Geo-lock' key as illustrated above encapsulates the main idea. }\label{fig:1d}
    \end{center}
\end{figure}

  The notion of QLV first appeared in the scientific literature in 2010 as a means of securing  real-time  communications to a \emph{unique} spatial position \cite{mal1}\footnote{Using the direct transfer of quantum entangled pairs in line-of-sight conditions as an aid to `tagging' (verifying) an object's location  appeared in the patent literature in 2006 \cite{kent1}.}. In  \cite{mal1} it was pointed out how by monitoring the location of a device using classical radio communications and  quantum information  embedded in the device, real-time data could be delivered to a specific geographic location. Data  would be sent to the receiver only whilst the QLV was validated. Through its use of classical communications, QLV can be made operational under most channel conditions including those  without a direct line-of-sight (LoS) component.
  Pivotal to this paradigm of  communication throughput conditioned on the location of the receiver is the anti-spoofing attributes of the specific QLV protocol (spoofing is always possible in classical-only systems \cite{chan}). In general terms, such anti-spoofing properties can be attributed to the quantum no-cloning theorem \cite{woot} coupled with the no-signaling principle of special relativity.  Since the work of \cite{mal1}  QLV has been further studied in numerous works with its deployment and security under a wide range of conditions and assumptions investigated. Details of this history and associated references can be found in \cite{malaney2} where the notion of QLV is discussed in the context of a quantum-enabled autonomous vehicle. In \cite{malaney2} arguments for QLV's security against any known attack are also presented.

  In this work we extend the notion of location-enabled communications discussed in \cite{mal1} to the decryption process itself. Using any pre-existing secret key (or any key, e.g. one generated dynamically with pre-existing shared entanglement) will not suffice for our purposes as it could be distributed to other adversaries at other locations (even though one  adversary is at the QLV validated location). However, as we show here, by using QDC and by obfuscating the quantum information needed by both the QDC and the QLV protocols until the time of the decryption and verification processes, we shall see how it is possible make the \emph{decryption} of data itself conditional upon the location of the receiver. Although more complex, and requiring some quantum technological developments (e.g. long-lived quantum memory), this form of  geo-encryption is free from the vulnerabilities that plague classical geo-encryption protocols.

\section{Quantum Geo-Encryption}
In quantum geo-encryption
  we  assume that most of the quantum information to be used is pre-stored in the device that will invoke the decryption (the decryptor).   This information could have been previously delivered to the decryptor via a quantum information transfer link such as fibre optic cable or LoS laser communications (or in real-time via teleportation). We will also assume that the stored quantum states are forced to be non-orthogonal, and in the wider context can take any allowable form, e.g. continuous variable (CV) states, qubits, qudits, hybrids, etc. We also allow for some states being entangled  - possibly with states at multiple reference stations (RSs). The RSs are at known locations, assumed to be legitimate members of the wider communication system, and assumed to have secure quantum and classical (e.g. via QKD) communications between themselves.

   The basic idea in quantum geo-encryption is that encoded signals (here we will take these signals to be classical)  sent by the RSs and arriving simultaneously at the receiver, instruct the decryptor what specific quantum states should be used (and for what process), any information (e.g. random unitary matrices) needed to remove any non-orthogonality introduced, any decoding instructions, and instructions on what output to produce (e.g. a classical output to be broadcast or a state to be teleported (or directly transferred) back to a specific RS). By measuring the round-trip time associated with these procedures the decryptor's location can be verified (see \cite{malaney2} for more details on QLV).

 Due to the wide variety of forms that quantum information can take, and the many variations on how that information can be transferred, stored, manipulated, and/or retrieved from the decryptor; unlimited versions of the quantum geo-encryption protocol are possible. For the purpose of focus, we will concentrate here on just one implementation of a geo-encryption protocol involving qubits, Gaussian states, and hybrids formed from combinations of such states. We choose this version simply for its ease in demonstrating the most important concepts of a quantum geo-encryption application.

   In terms of the annihilation and creation operators $\hat a,\,{\hat a^\dag }$,  the quadrature operators $\hat q,\hat p$  defined for photon states are ($\hbar=2$ assumed)
$\hat q = \hat a + \,{\hat a^\dag }$ and $\hat p = i({\hat a^\dag } - \hat a\,)$,
  satisfying  $\left[ {\hat q,\,\hat p} \right] = 2i$.
 The quadrature operators for a CV state with $N$ modes can be defined by the vector
%\begin{eqnarray}\label{p2}
${\bm{\hat R}_{1, \ldots ,n}} = \left( {{{\hat q}_1},\,{{\hat p}_1}, \ldots ,{{\hat q}_N},{{\hat p}_N}\,} \right)$.
%\end{eqnarray}
%Similarly, ${R_{1, \ldots ,n}} = \left( {{q_1},\,{p_1}, \ldots ,{q_n},{p_n}\,} \right)$ is defined for the real variables $q, p$ - the eigenvalues of the quadrature variables.
Gaussian CV  states are one of the work-horses of quantum information (e.g. \cite{gauss} for review).
An important and widely used Gaussian state, and one that will participate in our protocol,  is the coherent state $\left| \alpha  \right\rangle  = \exp \left( { - \frac{{{{\left| \alpha  \right|}^2}}}{2}} \right)\sum\limits_{n = 0}^\infty  {\frac{{{\alpha ^n}}}{{\sqrt {n!} }}\left| n \right\rangle }$ written here in terms of the Fock (number) states ${\left| n \right\rangle }$.
 Gaussian states are characterized solely by the first moments  $\left\langle {{{ \bm{\hat R}}_{1, \ldots ,N}}} \right\rangle $ and  for the $N=2$ case a covariance matrix  $\bm{M}$, which can be written
  $$\bm {M_s} = \left( {\begin{array}{*{20}{c}}
   \bm A & \bm C  \\
   {\bm{C}^T} & \bm{B}  \\
\end{array}} \right)$$
where
$\bm{A} = \tilde{a}{\bm{I_2}}\,,\,\bm{B} = \tilde{b}{\bm{I_2}}\,,\,\bm{C} = diag\left ( {{c_ + },{c_ - }} \right )$,
 $\tilde{a},\tilde{b},{c_ + },{c_ - } \in \mathbb{R}$, and
 ${\bm{I_2}}$ is the $2 \times 2$ identity matrix.
%The diagonal elements $a$ and $b$ are nonnegative real scalars and ${c_ + },{c_ - }$ are real scalars.
In this form the symplectic spectrum of the partially transposed covariance matrix is
$
{\nu _ \pm } = ( [ {{{\Delta  \pm \sqrt {{\Delta ^2} - 4\det \bm{M_s} } }} }]/ 2 )^{1/2}$ ,
%\end{eqnarray}
where $\Delta = \det \bm A + \det \bm B - 2\det \bm C$.
From the symplectic spectrum many fundamental properties of Gaussian states can be derived (see \cite{gauss}).
An important Gaussian state  is the two-mode squeezed vacuum (TMSV) state,
%also known as the Einstein-Podolski-Rosen (EPR) state.
described for two modes $a$ and $b$ as
$$\left| s \right\rangle  = \sqrt {1 - {\lambda ^2}} {\sum\limits_{n = 0}^\infty  {\left( { - \lambda } \right)} ^n}{\left| n \right\rangle _a}{\left| n \right\rangle _b},$$ where $\lambda  = \tanh (r) \in \left[ {0,1} \right],$ and where
${\left| n \right\rangle _a}$ and ${\left| n \right\rangle _b}$ are Fock  states of modes $a$ and $b$, respectively. Here,  $r$  is a   parameter quantifying the two-mode squeezing operator ${S_2}(r) = \exp \left[ {r\left( {\hat a\hat b - {{\hat a}^\dag }{{\hat b}^\dag }} \right)/2} \right]$.
%where ${\hat a}$ and ${\hat b}$ are the annihilation operators of the two modes.
The covariance matrix for the TMSV state can then be written
%can then be simply derived  in terms of  $v = \cosh (2r)$, the noise variance in the quadratures.
$${\bm{M_{T}}} = \left( {\begin{array}{*{20}{c}}
   {v{\bf{I}}} & {\left( {\sqrt {{v^2} - 1} } \right){\bf{Z}}}  \\
   {\left( {\sqrt {{v^2} - 1} } \right){\bf{Z}}} & {v{\bf{I}}}  \\
\end{array}} \right)$$
where the quadrature variance $v = \cosh (2r)$, and ${\bf{Z}}$~: = diag(1,-1). In wider geo-encryption protocols, TMSV states will be deployed partly for their use in holding entanglement between CV states at the RS  and CV (or hybrid) states at the decryptor.

In addition to Gaussian CV states we will also utilize discrete two-dimensional  states of the generic qubit form $a'\left| 0 \right\rangle  + b'\left| 1 \right\rangle$, ${\left| {a'} \right|^2} + {\left| {b'} \right|^2} = 1$,
which can be entangled together to form higher multi-dimensional states of various levels of entanglement. We will also utilize combinations such as $\left| \alpha  \right\rangle  \otimes \frac{1}{{\sqrt 2 }}\left( {\left| 0 \right\rangle  + \left| 1 \right\rangle } \right)$ to form
hybrid forms of such states such as $\frac{1}{{\sqrt 2 }}\left( {\left| {\alpha {e^{ - i\theta }}} \right\rangle \left| 0 \right\rangle  + \left| {\alpha {e^{i\theta }}} \right\rangle \left| 1 \right\rangle } \right)$ for some arbitrary phase $\theta$ (such states are not an absolute requirement for geo-encryption - but they will help illustrate its operation better).

 In QDC  \cite{ping,ping2,ping3,ping4} we negate the use of QKD, but instead encrypt the message we wish to send directly into a series of quantum states (see \cite{QDC1} for recent review). Here, we will utilize a form of QDC based on entangled qubit pairs, namely the so-called ping-pong protocol \cite{ping}.\footnote{We use this  version of QDC  here simply to illustrate  the quantum  geo-encryption paradigm. We could also use other variants of QDC  that have better communications-security attributes, that use states other than qubits, and that sometimes invoke additional classical communications.} In the ping-pong protocol a maximally  entangled qubit pair, say $\left| \psi^+  \right\rangle  = \frac{1}{{\sqrt 2 }}\left( {\left| 0 \right\rangle \left| 1 \right\rangle  + \left| 1 \right\rangle \left| 0 \right\rangle } \right)$,
  is shared between the sender and receiver. Depending on whether a 0 or a 1 is to be encoded, the sender applies to its qubit an $I = \left( {\left| 0 \right\rangle \left| 0 \right\rangle  + \left| 1 \right\rangle \left| 1 \right\rangle } \right)$ or an $\sigma_z = \left( {\left| 0 \right\rangle \left| 0 \right\rangle  - \left| 1 \right\rangle \left| 1 \right\rangle } \right)$
 operation, respectively. This qubit (the travel qubit) is sent to the receiver, who then performs a Bell measurement on both qubits. From the possible two results of this measurement, $\left| \psi^+  \right\rangle  = \frac{1}{{\sqrt 2 }}\left( {\left| 0 \right\rangle \left| 1 \right\rangle  + \left| 1 \right\rangle \left| 0 \right\rangle } \right)$ or $\left| \psi^-  \right\rangle  = \frac{1}{{\sqrt 2 }}\left( {\left| 0 \right\rangle \left| 1 \right\rangle  - \left| 1 \right\rangle \left| 0 \right\rangle } \right)$, the encoded message is inferred. During this process the sender-receiver pair can do other tests to monitor for the presence of an eavesdropper.
  In our specific form of the geo-encryption protocol we assume the (travel) qubit sent by the sender to the receiver, is teleported to the receiver using another pair of maximally entangled qubits that pre-exist between the sender and the receiver. The classical radio communication associated with this teleportation will form part of the  geo-encryption protocol.

  \begin{figure}[!t]
    \begin{center}
    {\includegraphics[width=3.4 in]{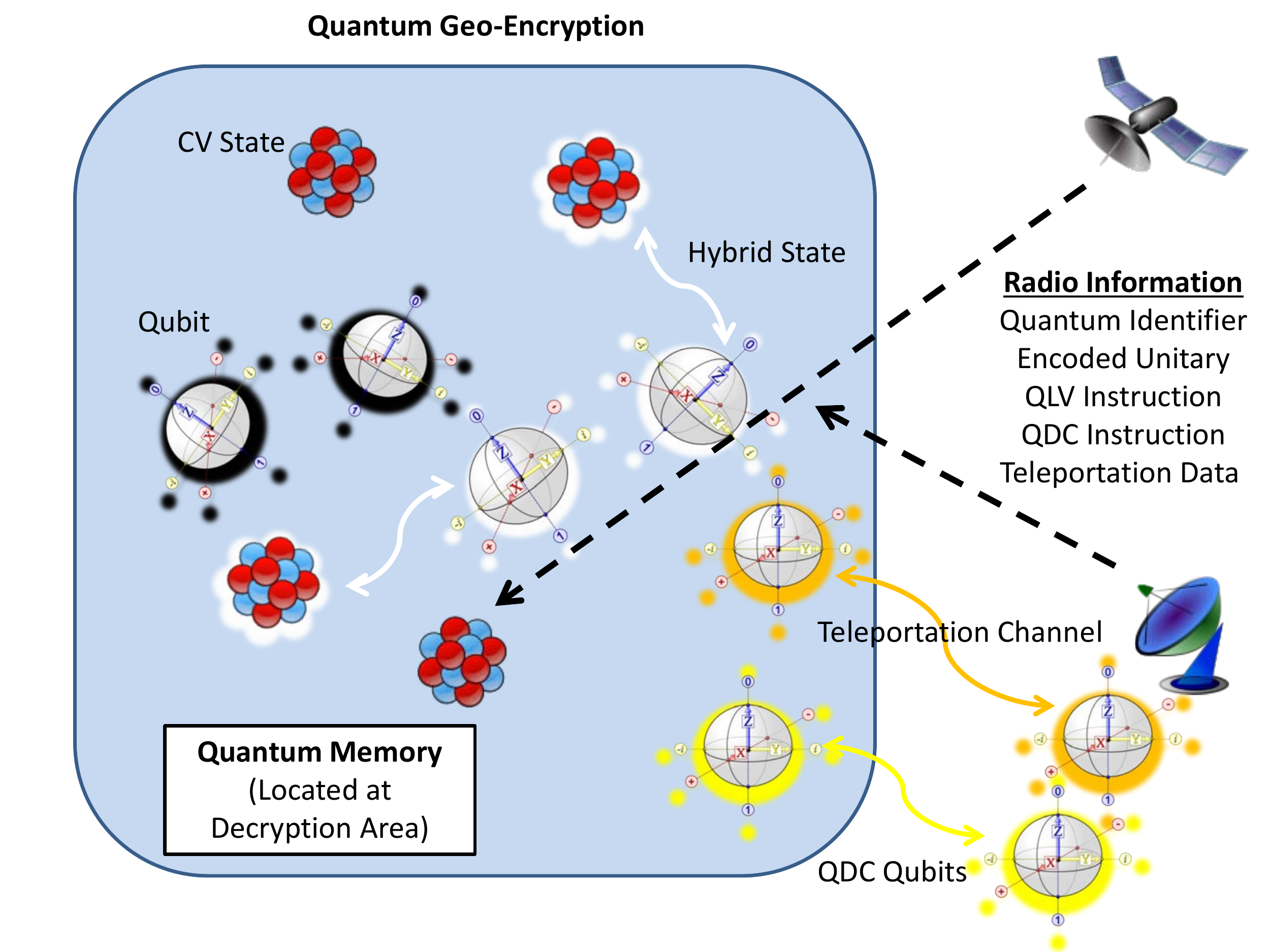}}
    \caption{Quantum geo-encryption using quantum information embedded at the decryptor. The decryption instructions needed to access and act on the relevant quantum information are sent from a collection of satellite and/or terrestrial transmitters via classical radio communications. Prior to receiving this information it is impossible for the decryptor to ascertain which quantum states are associated with the QDC process and which states are associated with the QLV process. Different shadowed colors indicate different roles played by the qubits. Wiggled-arrows indicate entanglement. Dependent on the instructions received at the decryptor, some classical result is sent by the decryptor back to the radio transmitters (who now act as receivers).  }\label{fig:1c}
    \end{center}
\end{figure}

  In our geo-encryption protocol we will obfuscate all quantum information to be used  in several ways. First, all stored information used in the wider geo-encryption protocol, including all qubits, will have  random unitary transformations applied to each state prior to being transferred to the decryptor. The classical information describing each of these unitaries will be  held by the legitimate system in an encoded and distributed form  - and will also form part of the radio communications of the geo-encryption protocol.\footnote{We will assume all classical signals are encoded with some  block-size (which can be large and made \emph{a priori} unknown)  and signals from multiple RSs are needed for decoding. The number of RSs  used (randomly selected from those available) can form part of the decoding instructions, e.g. coding the block-size used.} Some qubits will be part of the QLV protocol and some will be part of the QDC protocol. Some qubits will be entangled with other qubits and some qubits will be entangled with CV states. Some of the entangled states held by the decryptor will be entangled with states still held by the legitimate system. The QLV protocol will also partially use states from this same qubit grouping - the idea being that since the receiver does not know which states are to be used in each protocol - and since the QLV protocol demands the states it needs be at the correct location - then the QDC protocol is also tied to states that must be at the same location.  Any attempt to move quantum information that is required by the QDC protocol (to another device at another location) will inevitably lead to a violation of the QLV protocol. By checking simultaneously that the QLV passes all verification tests, we can ensure the decryption process occurs  only at the required location.

  Since we can also dictate the time when the required classical information sent by the RSs reach the decryptor, we can also ensure the decryption occurs at a specific time (at least not before a specific time). Avoiding delayed decryption of a complete message stream would require the QDC and QLV outputs be intertwined (e.g. embed some of the information needed by the QLV protocol within the QDC protocol).\footnote{That is, strictly speaking we have thus far only considered the case where decryption occurs in real-time (i.e. as soon as the decryptor has the required information). However, it is straightforward to avoid delayed encryption by ensuring that (at least some) information required for QLV is encoded in the QDC quantum states - and that this information is not available until all states that can be currently decrypted are in fact decrypted.}

   A schematic of a geo-encryption protocol is given in Fig.~2. Not explicitly shown in Fig.2 are other quantum states (possibly entangled) and other entanglement between quantum states stored in memory and states stored at the RSs. Also not shown are quantum transmitters (receivers) that may be added  so as to transmit (receive) quantum information to (from) the decryptor if dictated to do so by the QLV protocol when conditions allow.

Using other quantum states held by the decryptor designated for tasks beyond QDC and QLV, straightforward extensions of the geo-encryption protocol can be used to produce any quantum task (or output). For example, quantum computing steps could be forced to occur at a specific computing chip at a specific location. Such quantum computation steps could be part of  wider distributed quantum algorithm thereby adding a new layer of security to distributed computing not offered by classical machines. Routing of information - classical and/or quantum - could also be enhanced in the sense that routing decisions could be enforced to be viable only at certain location-specific routers. This would allow for a chosen geographic path to be chosen as the sole route for information transfer. Clearly, many other enhanced application scenarios exist for the geo-encryption paradigm.

Quantum geo-encryption is a rather more complex beast than its classical counterpart, and a series technological advances are required before its real-world implementation can occur. However, in the simplest forms of quantum geo-encryption the main technological advance needed is the development of reliable and long-lived quantum memory. Rapid progress is being made in this endeavor (e.g. see \cite{memory2, memory}), and as such it is not inconceivable that quantum geo-encryption will be deployable in the not-so-distant future.

\section{The Decrypting Region}

Thus far we have assumed the decryptor can be forced to invoke decryption at a unique  location - a specific point in geographical space. In reality, due to inevitable noise (and any processing delay) in the location verification procedures this is not actually true in practice - devices that appear close to (but not at)  the location can (and should)  pass the verification test. What occurs in practice is we limit the decryption process to occur in an  \emph{effective region} or area of geographical space. The dimensions of this region can be thought of as being set according to a probability $P_c$ that the received timings are consistent with the decryptor being at the location it claims to be  at (the likelihood of the timing measurement is used by the QLV decision-making process). Setting a required $P_c$ can therefore be visualized as setting an effective region in space (e.g. an error ellipse) within which a position - determined from the timings  - must lie within with probability $P_c$. The probability of any adversary spoofing the system approaches zero the further she actually is from this region.

 The size of the effective region is highly dependent on the nature of the radio communications channel used in the quantum geo-encryption protocol, most importantly on the weight of the non-LoS component in the channel relative to  scattered components in the channel. Note that non-LoS components can be of a real or virtual nature. In the latter case, no visible LoS between the sender-receiver pair exists, but energy in the LoS component is still detected by the receiver. Such an effect is due to the well-known penetration power (dependent on the radio frequency and the composition of the obstructing object) of radio waves. Such penetration capabilities underpin  so-called `see-through-the wall' radars \cite{loc3} and Ultra-Wide-Band (UWB) location systems \cite{loc1,loc2}. Recent advanced deployments have demonstrated virtual LoS detection through 20cm thick concrete walls \cite{loc3}.\footnote{Henceforth we will ignore the  tiny delay caused by propagation through objects - as this error (related to the dimension of the obstruction) can be encapsulated into the effective region (the set $P_c$ value). In the rare circumstance where no set-up can allow for a virtual LoS component to be present at the decryptor, the error introduced can be estimated and also folded into a new effective region. Details of such issues and strategies around optimal NLoS detection/mitigation are beyond the scope of the present work.}

To understand in general terms the scale of the decrypting region   let us consider a very general model for the wireless channel - the Rician channel. Such a channel between a sender $a$ with $N_a$ antennae (linear array)   and a receiver $b$ with one antenna can be described by $\mathbf{h}_{ab}$ (the $1\times N_a$ channel vector), where
$$ \mathbf{h}_{ab} =\sqrt{\frac{K_{ab}}{1+K_{ab}}}\mathbf{h}_{ab}^{o} + \sqrt{\frac{1}{1+K_{ab}}}\mathbf{h}_{ab}^{r},
$$
where $K_{ab}$ denotes the Rician $K$-factor of the channel, $\mathbf{h}_{ab}^{o}$ denotes the LoS component of the  channel, and $\mathbf{h}_{ab}^{r}$ denotes the scattered component of the  channel - the elements of which are assumed to be i.i.d complex Gaussian random variables with zero mean and unit variance, i.e., $\mathbf{h}_{ab}^{r}\sim\mathcal{CN}\left(\mathbf{0}_{1\times N_a},\mathbf{I}_{N_a}\right)$. In the above, $\mathbf{h}_{ab}^{o}$ can be expressed as
$
\label{h_o} \mathbf{h}_{ab}^{o} = \begin{bmatrix}
  1,\cdots,\exp\left(j2\pi\left(N_a-1\right)\delta_a\cos\theta_{ab}\right)
\end{bmatrix},
$
where $\delta_a$ denotes the constant spacing (in wavelengths) between the antennas of the linear array, and $\theta_{ab}$ is the angle defined by the direction of the linear array and the direction to the intended receiver. In the following to simplify the discussion we will ignore the directional information afforded to us  by the linear array (i.e. we ignore angle-of-arrival information) - bearing in mind that its inclusion usually assists in the performance of any location-verification algorithm.

The critical determination to be made in any channel is the time-of-arrival (ToA) information obtained by the RSs. The best approach to use for this problem is strongly dependent on the signal-to-noise (SNR) ratio of the channel. In the high-noise SNR regime it is known that the Cramer Rao Bound (CRB) provides a tight bound on the timing error, whereas in the low-noise regime the Ziv-Zakai Bound (ZZB) is known to be tighter (more pragmatically useful) e.g. \cite{33}.

Consider some signal pulse $\Phi (t) = \omega s(t - T) + n(t)$, for some signal template $s()$, time $t$, channel coefficient $\omega$, ToA $T$, and additive white Gaussian noise $n(t)$ with zero mean and spectral density $N_0/2$. The CRB on the ToA is then given by ${\left( {2\sqrt 2 \pi \beta \sqrt {{\rm{SNR}}} } \right)^{ - 1}}$, where ${\beta ^2} = \left( {\int_{ - \infty }^\infty  {{f^2}\left| {S{{(f)}^2}} \right|df} } \right)/ \left( {\int_{ - \infty }^\infty  {\left| {S{{(f)}^2}} \right|df} } \right)$, and where $f$ is the frequency and $S(f)$ is the Fourier transform of $s(t)$. As we can see, in principal for large enough SNR the CRB can be driven to zero. This CRB result is modified in the presence of multipath components (i.e. some finite $K_{ab}$). Nonetheless, as long as $K_{ab}>0$, in principal the ToA can still be driven to zero.

The ZZB bound  has the advantage that \emph{a priori} channel information can be easily encapsulated to form a better estimator. In the high SNR limit and for unbiased conditions the ZBB provides the same bounds as the CRB. The optimal detector for the ZZB ToA estimator has been derived in \cite{woe} for UWB Rician channels. We note the ZBB is based on constructions of a likelihood ratio test (LRT) for optimum decision rules. Once having determined the timing error associated with the relevant ToA its role in setting the scale of the decrypting region can be determined via analysis.

For ease of exposition, here we will determine a CRB estimate of the location error bounds given some Gaussian distribution on the ToA timing errors (e.g. \cite {patwari, malanlog, chenxi1}).
In the following we denote the true location of the decryptor by ${\zeta}_{0} = \left[{x}_0,{y}_0\right]$ and the location of the $n^{\text{th}}$ RS by $\zeta_n = \left[x_n,y_n\right]$.
%We assume that the ToA timing  at the $n^{\text{th}}$ RS follows the normal distribution.
 To counteract the impact of unknown bias in the timings (e.g. synchronization offsets or constant NLoS bias) it is normally useful to move to a time difference of arrival setting (TDoA).  To achieve this we denote the time difference relative to that measured by RS $1$ at the $n^{th}$ RS as $\phi_n$, then the logarithm of the distribution of $\phi_n$ is obtained as
$$
 -\ln f\left(\phi_n\right) = \dfrac{\left(\phi_n-\frac{d_n-d_1}{c}\right)^2}{4c^2\sigma_t^2},
$$
where $c$ is the speed of light, $\sigma_t^2$ is the variance of the timings, and the distance between the $n^{\text{th}}$ RS and the decryptor is expressed as
$
 d_n = \sqrt{\left(x_n-{x}_0\right)^2 + \left(y_n-{y}_0\right)^2}.
$
The Fisher matrix for the TDoA scheme can then be written
$$ \mathbf{J}\left({\phi_n}\right) = \left[ {\begin{array}{*{20}{c}}
   {{J\left({\phi_n}\right)_{11}}} & {{J\left({\phi_n}\right)_{12}}}  \\
   {J\left({\phi_n}\right)_{21}} & {{J\left({\phi_n}\right)_{22}}}  \\
\end{array}} \right],
$$
where
$$J\left({\phi_n}\right)_{11}=\dfrac{1}{2c^2\sigma_t^2}\sum_{n=2}^N\left(\cos\theta_n-\cos\theta_1\right)^2,
$$
%b\sum_{n=1}^N\dfrac{\cos^2\theta_n}{d_n} +
$$ J\left({\phi_n}\right)_{22}
=\dfrac{1}{2c^2\sigma_t^2}\sum_{n=2}^N\left(\sin\theta_n-\sin\theta_1\right)^2,
$$
%b\sum_{n=1}^N\dfrac{\sin^2\theta_n}{d_n} +
and
%$$ J\left({\phi_n}\right)_{12} = J\left({\phi_n,\varphi_n}\right)_{21}
%= \dfrac{1}{{2c^2\sigma_t^2}}\sum_{n=2}^N\left(\sin\theta_n-\sin\theta_1\right)\left(\cos\theta_n-\cos\theta_1\right).
$$
\begin{array}{l}
 J{\left( {{\phi _n}} \right)_{12}} =
 \frac{1}{{2{c^2}\sigma _t^2}}\sum\limits_{n = 2}^N {\left( {\sin {\theta _n} - \sin {\theta _1}} \right)} \left( {\cos {\theta _n} - \cos {\theta _1}} \right) \\
 \end{array}.$$
Here,
$
\theta_n = \arctan[({y_n-y_0})/({x_n-x_0})].
$
We can now express the covariance matrix of the decryptor's location as
$
\mathbf{V}_{pos} = \mathbf{J}^{-1}.
$
%where $\mathbf{J}\in\left\{\mathbf{J}\left({\phi_n}\right)\right\}$.
We further define $\mathbf{V}_{pos}$ as
$$ \mathbf{V}_{pos} = \left[
\begin{array}{*{20}{c}}
  \sigma_{x}^2&\sigma_{xy}\\
  \sigma_{yx}&\sigma_{y}^2
\end{array}\right],
$$
where $\sigma_{xy}=\sigma_{yx}$. We denote the estimated decryptor's location by ${\zeta}_e = \left[{x}_e,{y}_e\right]$, and the correlation coefficient by
$
 \rho = {\sigma_{xy}}/({\sigma_x\sigma_y}).$
Making the usual assumption that the likelihood function of the parameters to be determined is Gaussian in the vicinity of the true values, the distribution of the estimated decryptor's location can be expressed as

%
%$$P({\zeta _e}) = \frac{1}{{2\pi \sqrt {1 - {\rho ^2}} {\sigma _x}{\sigma _y}}}\exp \left\{ \begin{array}{l}
%  - \frac{1}{{2\left( {1 - {\rho ^2}} \right)}} \\
% \left( \begin{array}{l}
% \frac{{{{\left( {{x_e} - {x_0}} \right)}^2}}}{{\sigma _x^2}} + \frac{{{{\left( {{y_e} - {y_0}} \right)}^2}}}{{\sigma _y^2}} \\
%  - 2\frac{{\rho \left( {{x_e} - {x_0}} \right)\left( {{y_e} - {y_0}} \right)}}{{{\sigma _x}{\sigma _y}}} \\
% \end{array} \right) \\
% \end{array} \right\}.$$

$$\begin{array}{c}
 P({\zeta _e}) = \frac{1}{{2\pi \sqrt {1 - {\rho ^2}} {\sigma _x}{\sigma _y}}} \times  \\
 \exp \left\{ \begin{array}{l}
  - \frac{1}{{2\left( {1 - {\rho ^2}} \right)}} \\
 \left( \begin{array}{l}
 \frac{{{{\left( {{x_e} - {x_0}} \right)}^2}}}{{\sigma _x^2}} + \frac{{{{\left( {{y_e} - {y_0}} \right)}^2}}}{{\sigma _y^2}} \\
  - 2\frac{{\rho \left( {{x_e} - {x_0}} \right)\left( {{y_e} - {y_0}} \right)}}{{{\sigma _x}{\sigma _y}}} \\
 \end{array} \right) \\
 \end{array} \right\}. \\
 \end{array}$$

\begin{figure}[!t]
    \begin{center}
    {\includegraphics[width=3.4 in]{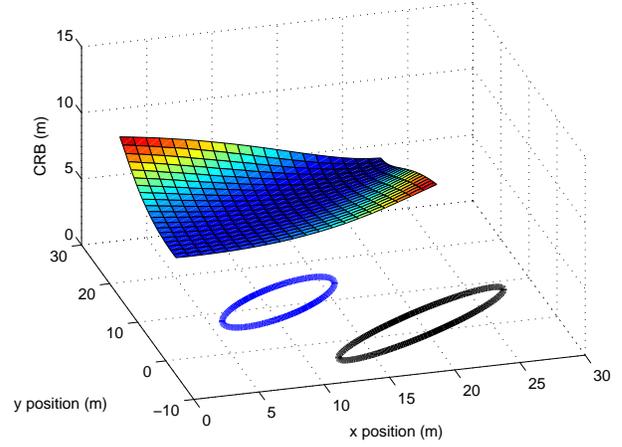}}
    \caption{The Cramer Rao Bounds  associated with a typical set-up of quantum geo-encryption. Here we have used anticipated timing errors associated with state-of-the-art radio detection systems and have used a TDoA scheme using four radio receivers. We have also indicated the associated error ellipses at two points in the simulated space which illustrate the effect noise can have in blurring the allowed location into an effective allowed region. Here, we have assumed processing delays are negligible.}

    \end{center}
\end{figure}

Using the above analysis, Fig.~3 displays the anticipated CRB's for a  set-up of four RSs. Here we have assumed  $c \sigma_t  = 1.8$m and the RSs are  of order 100m away toward the corners of the simulation area (moving the RSs beyond about 1km has the simple effect of flattening the CRB plane to a constant error of about 7m). As we can see very reasonable errors in the location verification will be possible for such settings. We have also shown the error ellipse for two specific points (the ellipse center indicates the point) using an eigenvalue analysis of the inverse Fisher matrix to set the scale and orientation of the ellipse axes. Multiplying these axes by  a factor of three will lead to a scenario in which a location estimate of the decryptor (which if truly  at the ellipse center) would be placed outside the expanded ellipse with    probability $\epsilon= 0.01$.
%More generally, this probability will approach zero - as dictated $P({\zeta _e})$ above -  as the size of the allowed region is expanded further.

 A more sophisticated analysis of the decrypting region can be given, as well as a more detailed probability analysis, that takes into account all \emph{a priori} information and allows for decision-making even when a location cannot be determined - for example when less than three RSs with accurate ToAs are available (e.g. see \cite {malaney2} where a more direct decision-making process for QLV was adopted using  optimal decision theory derived from a LRT). Nonetheless, the analysis  provided here already shows
that for anticipated conditions a well-defined and pragmatically-useful decryption region can be readily determined. For a given $\epsilon$, the size of the region can be very small (of order of cm) or very large (of order of km), dependent on many factors such as timing resolution, frequency bandwidth, transmitter power (regulatory limits etc) and channel conditions. As such, quantum geo-encryption can in principle be used in a wide range of real-world scenarios.

 \section{Conclusions}
 In this work we have introduced quantum geo-encryption - a protocol that can allow data-decryption to occur at only one location and one specified time. We have also discussed how this same protocol can be extended to provide the same spatial and temporal characteristics to  any quantum task. Other useful security outcomes, such as enforcing  a pre-specified route for data transfer, are also possible.
 In reality, issues arise when using an actual specific \emph{point} for the allowed location.
In real-world deployments the allowed location in practice becomes an allowed effective region. This region accommodates noise issues that allow for timings that are not exactly consistent with a single geographical point. However, when assessment of the channel conditions is taken, the size of such an effective region can be readily determined and the probability of spoofing the system  assessed.
% By increasing the size of the allowed region this probability of spoofing can be driven arbitrarily close to zero - an outcome that is true even when non-LoS conditions prevail in some of the channels.

 The specific protocol we have used in this work for the geo-encryption can be replaced by many different variants, each having their own trade-offs in terms of complexity versus performance.
 %Many variations in the QLV process exists as do many variation in the QDC process.
 Quantum geo-encryption opens up, for the first time, the possibility of delivering encrypted communications  in the knowledge that the received signals can only be decrypted by a device at a unique geographical location (free from the type of attacks that have hindered deployment of classical geo-encryption). In the coming years, the  technological innovations required  for a  quantum-geo encryption implementation  (e.g. long-lived quantum memory) will be at hand.
 %A wide range of security scenarios can be considered for its use.

%
%%
%\section*{Acknowledgment}
%% optional entry into table of contents (if used)
%%\addcontentsline{toc}{section}{Acknowledgment}
%This work has been supported by a University of New South Wales Research Award.

 %(unconstrained entanglement and the ability to conduct an indefinite number quantum of operations at no time penalty).

%\subsection{Preliminaries}

%

%
%
%\appendix

\end{document}